\documentclass[10pt]{article}

\usepackage{basecompute}

\renewcommand{\basecomputelogo}{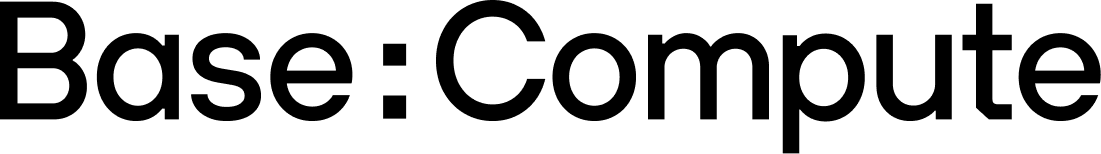}

\basetitle{BaseRT: Advancing Best-in-Class LLM Inference\\ with Apple M5 Neural Accelerators}

\author{
  Fabian Waschkowski\thanks{Correspondence: \href{mailto:fabian@basecompute.co}{fabian@basecompute.co}} \quad Prabod Rathnayaka \quad Lukas Wesemann \\[8pt]
  \textit{Base Compute, Melbourne, Australia} \\
}

\date{}

\begin{document}

\maketitle

\begin{abstract}
Apple's M5 generation introduces a redesigned GPU architecture in which every core carries a dedicated Neural Accelerator: on-die matrix units exposed through the Metal~4 tensor API. We show that BaseRT, our native Metal inference runtime for large language models on Apple Silicon, exploits these units to push inference throughput on Apple hardware substantially beyond both llama.cpp and MLX. Building on BaseRT's framework-free design, we add a family of hand-written Metal~4 tensor-core kernels (including dense and mixture-of-experts GEMM and flash-attention prefill kernels) that route the compute-bound matrix multiplications of inference through the M5 Neural Accelerators while leaving the memory-bound decode path on our existing specialised kernels. On an Apple M5 Pro, across fifteen model configurations spanning the Qwen3, Qwen3.5/3.6, Llama~3.2, and Gemma~4 families from sub-1B to 35B parameters, BaseRT delivers up to $6.4\times$ higher prompt-processing throughput than llama.cpp and $3.9\times$ higher than MLX, with the largest margins on the mixture-of-experts models where matrix multiplication dominates, while maintaining its lead on decode of up to $1.75\times$ over llama.cpp and $1.33\times$ over MLX. These results establish a new performance ceiling for on-device LLM inference and show that the M5's tensor cores are the decisive lever for prompt processing on Apple Silicon. BaseRT is publicly available at \url{https://github.com/basecompute/baseRT}.
\end{abstract}

\section{Introduction}

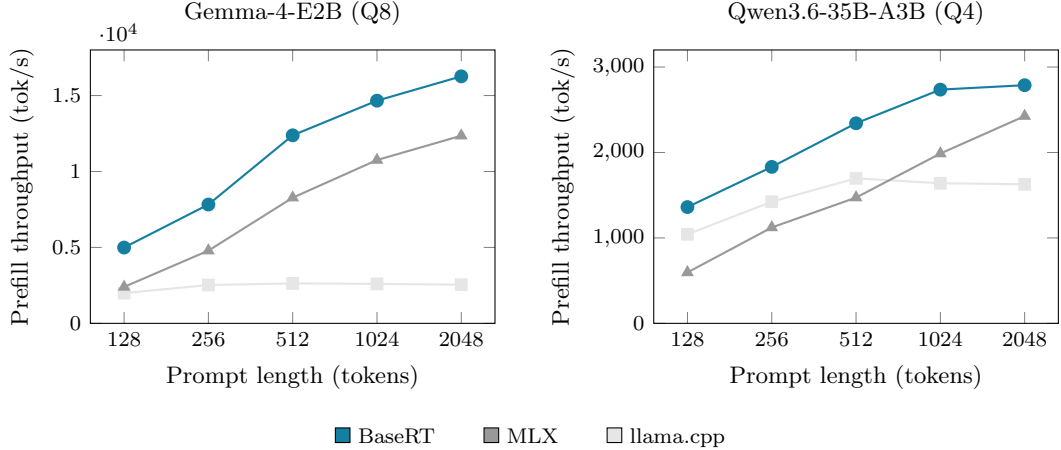
\begin{figure}[t]
  \centering
  \begin{tikzpicture}
    \begin{groupplot}[
      group style={group size=2 by 1, horizontal sep=2.1cm},
      width=0.42\textwidth, height=5.2cm,
      xmode=log, log basis x=2,
      xtick={128,256,512,1024,2048},
      xticklabels={128,256,512,1024,2048},
      xlabel={Prompt length (tokens)},
      ylabel={Prefill throughput (tok/s)},
      xlabel style={font=\small},
      ylabel style={font=\small},
      xticklabel style={font=\footnotesize},
      yticklabel style={font=\footnotesize},
      title style={font=\small},
      every axis plot/.append style={thick, mark options={scale=1.1, fill opacity=1}},
    ]
    \nextgroupplot[ymin=0, ymax=18000, title={Gemma-4-E2B (Q8)}]
      \addplot[gray!20, mark=square*, mark options={fill=gray!20}] coordinates {
        (128,1992.3) (256,2524.2) (512,2631.2) (1024,2599.5) (2048,2546.6)};
      \addplot[black!40, mark=triangle*, mark options={fill=black!40}] coordinates {
        (128,2397.6) (256,4787.8) (512,8273.9) (1024,10752.3) (2048,12354.9)};
      \addplot[basecompute, mark=*] coordinates {
        (128,4991.2) (256,7824.6) (512,12380.2) (1024,14660.7) (2048,16263.9)};
    \nextgroupplot[ymin=0, ymax=3200, title={Qwen3.6-35B-A3B (Q4)}]
      \addplot[gray!20, mark=square*, mark options={fill=gray!20}] coordinates {
        (128,1042.4) (256,1424.0) (512,1697.5) (1024,1639.7) (2048,1628.1)};
      \addplot[black!40, mark=triangle*, mark options={fill=black!40}] coordinates {
        (128,595.9) (256,1122.3) (512,1472.2) (1024,1987.6) (2048,2425.8)};
      \addplot[basecompute, mark=*] coordinates {
        (128,1361.1) (256,1831.4) (512,2342.1) (1024,2736.0) (2048,2787.6)};
    \end{groupplot}
  \end{tikzpicture}
  \par\vspace{0.5em}
  {\footnotesize
    \tikz[baseline=-0.8ex]{\filldraw[fill=basecompute,draw=black] (0,-0.8ex) rectangle (1.6ex,0.8ex);}~BaseRT\hspace{1.8em}
    \tikz[baseline=-0.8ex]{\filldraw[fill=black!40,draw=black] (0,-0.8ex) rectangle (1.6ex,0.8ex);}~MLX\hspace{1.8em}
    \tikz[baseline=-0.8ex]{\filldraw[fill=gray!20,draw=black] (0,-0.8ex) rectangle (1.6ex,0.8ex);}~llama.cpp}
  \caption{BaseRT delivers best-in-class prompt-processing (prefill) throughput on Apple M5 Pro by utilising the chip's per-core GPU Neural Accelerators. Left: Gemma-4-E2B (8-bit), where BaseRT leads llama.cpp by up to $6.4\times$ and MLX by up to $2.1\times$ as the prompt grows to 2048 tokens. Right: Qwen3.6-35B-A3B (4-bit), a mixture-of-experts model, where BaseRT leads both baselines at every prompt length.}
  \label{fig:hero}
\end{figure}

The case for on-device inference, i.e.\ running large language models locally on consumer and professional hardware rather than in the cloud, has grown steadily stronger. Global inference compute is projected to overtake training compute for the first time in 2027, and AI inference is expected to account for $40\%$ of global data-center compute by 2030~\cite{mckinsey2025aiworkloads}, placing increasing pressure on the economics of cloud-only serving. Four independent drivers push inference toward local devices. \textit{Privacy and data residency}: prompts processed in the cloud are subject to multi-tenancy and cross-border routing risk, which Gartner projects will cause over $40\%$ of AI-related data breaches by 2027~\cite{gartner2025crossborder, nist2024genai}. Local inference keeps data within controlled, air-gapped environments. \textit{Latency}: on-device inference avoids network, queueing, and scheduling overhead, enabling sub-100\,ms interactive responses for real-time robotic~\cite{huang2025corki} and agentic workloads~\cite{masterman2026agentic,jang2025edgefirst}. \textit{Connectivity}: local execution removes the hard dependency on external network and shared cloud infrastructure whose failures repeatedly render AI systems unavailable~\cite{anthropicstatus2026, openaistatus2026}. \textit{Token costs}: local inference converts linear per-token pricing into a fixed hardware cost with marginal cost approaching zero at scale~\cite{mckinsey2025neweconomics}. The appetite for cost-efficient open models is already visible, as roughly a third of all tokens on OpenRouter now go to open-weight models~\cite{mozilla2026opensource}.

Among edge platforms, Apple Silicon stands out for its combination of high memory bandwidth, large unified memory capacity, and a mature GPU compute stack~\cite{benazir2025profiling}. Its unified memory architecture removed the traditional CPU--GPU memory segregation, giving the GPU direct high-bandwidth access to full system memory and making it practical to run quantised models locally at interactive speeds~\cite{stanfordaiindex2025}. In prior work we introduced BaseRT~\cite{basert2026}, an LLM inference runtime built directly on Apple's Metal GPU API with no intermediate framework, which through chip-specific kernel fusion, unified-memory-aware data layout, and low-overhead dispatch achieved the highest reported inference throughput on Apple Silicon, ahead of both the widely deployed llama.cpp~\cite{llamacpp} and Apple's purpose-built MLX~\cite{mlx} framework.

The arrival of the Apple M5 architecture changes the optimisation landscape. Where earlier M-series GPUs executed matrix multiplication (matmul) on general-purpose SIMD units, the M5 introduces a redesigned GPU in which every core carries a dedicated Neural Accelerator: an on-die matrix engine, analogous to the tensor cores of discrete GPUs, purpose-built for the dense matmuls at the heart of transformer inference~\cite{apple2025m5}. Apple exposes these units to third-party developers through the Metal~4 tensor API and its Metal Performance Primitives cooperative-tensor operations~\cite{apple_metal4, apple_mpp, apple_wwdc25_tensor}. Driving these units requires kernels written directly against the new tensor API, and a runtime's performance on the compute-bound phases of inference depends on how well the accelerators are utilised.

This paper shows that BaseRT's framework-free architecture is well suited to exploiting this new hardware. We extend BaseRT with a family of hand-written Metal~4 tensor-core kernels for the compute-bound matmuls of inference: dense and mixture-of-experts (MoE) GEMM, fused expert projections, and the QK$^\top$ and PV products of prefill attention. These kernels route that work through the M5 Neural Accelerators, while leaving the memory-bandwidth-bound decode path on our existing specialised kernels where the tensor units offer no benefit. The result is a sharp widening of BaseRT's lead over both baselines, concentrated exactly where the workload is compute-bound: prompt processing and MoE inference (Figure~\ref{fig:hero}).

\subsection{Contributions}
Our contributions are as follows:
\begin{itemize}
    \item A family of hand-written Metal~4 cooperative-tensor kernels that drive the Apple M5's per-core GPU Neural Accelerators, covering dense GEMM, MoE expert GEMM, fused expert projections, and prefill attention.
    \item A workload-aware dispatch strategy that applies the tensor-core path only to the compute-bound phases of inference, preserving BaseRT's existing low-overhead kernels for memory-bound decode, with a clean fallback to the SIMD path on pre-M5 hardware.
    \item A comprehensive evaluation on the Apple M5 Pro across fifteen model configurations, establishing new state-of-the-art throughput on Apple Silicon and quantifying where the tensor cores matter: up to $6.4\times$ prefill throughput over llama.cpp, with decode gains bounded by memory bandwidth.
\end{itemize}

\section{Background: The M5 GPU and Metal~4 Tensor Operations}

\subsection{From SIMD Matmul to Neural Accelerators}
Every Apple M-series GPU from the M1 onward supports \texttt{simdgroup\_matrix}, a set of Metal intrinsics that cooperatively compute small fixed-size matrix products across the 32 lanes of a SIMD group. This is the mechanism on which BaseRT's tiled GEMM and FlashAttention kernels were originally built. It runs on the GPU's general-purpose arithmetic units: matrix multiplication competes with every other instruction for the same execution ports.

The M5 breaks this coupling. Its redesigned GPU places a dedicated Neural Accelerator inside each GPU core: a hardware matrix-multiply-accumulate engine that is the Apple Silicon analogue of a discrete GPU's tensor core, delivering a large step-change in peak GPU compute for AI workloads relative to the M4~\cite{apple2025m5}. Crucially, the M5 continues to advertise the \texttt{MTLGPUFamily.apple9} feature set, so the accelerators are not reached by querying a capability flag. They are reached only through a new programming interface.

\subsection{The Metal~4 Tensor API}
The new interface is Metal~4, which introduces a first-class \texttt{tensor} resource for multidimensional data, whose storage may live in device or threadgroup memory, and, through the Metal Performance Primitives (MPP) library, a set of \texttt{mpp::tensor\_ops} primitives, most importantly a \texttt{matmul2d}, issued collectively at simdgroup or threadgroup scope to drive the Neural Accelerators~\cite{apple_metal4, apple_mpp, apple_wwdc25_tensor}. A kernel declares tensor views over its operands, requests a destination cooperative tensor, i.e.\ a tensor whose elements are partitioned across the thread-local storage of the participating threads, and accumulates a sequence of tile products into it. Operands may stream directly from device memory or be staged in threadgroup memory. The cooperative-tensor accumulator remains resident across the reduction loop. Notably, threadgroup-memory staging is not necessary for peak GEMM on Apple Silicon: kernels are fastest reading operands straight from device memory and letting the cache hierarchy work~\cite{apple_mpp}. This is a materially different kernel structure from the \texttt{simdgroup\_matrix} model, and code must be compiled against the Metal~4 language revision (\texttt{-std=metal4.0}) to use it~\cite{apple_msl}. Because BaseRT hand-writes and owns every GPU kernel it dispatches, adopting this new path is a self-contained addition to its shader library.

\subsection{Utilising Tensor Cores}
LLM inference has two phases with opposite performance characteristics. Prefill (prompt processing) multiplies the prompt's activation matrix, one row per token, by the layers' weight matrices: because all tokens are processed together, arithmetic intensity grows with sequence length, making prefill compute-bound and dominated by these GEMMs and by the QK$^\top$ and PV attention products, whose cost scales with context. This is precisely the regime in which a dedicated matmul engine pays off. Decode generates one token at a time: each step multiplies a single activation vector against the weights ($M{=}1$ GEMV), streaming the entire weight set from memory for a handful of arithmetic operations per byte. Decode is memory-bandwidth-bound, and the M5's unified memory bandwidth, not its matmul throughput, is the binding constraint. A tensor core cannot accelerate a phase that is waiting on memory. This asymmetry, which we confirm empirically in Section~\ref{sec:eval}, is the organising principle of BaseRT's M5 design.

\section{Tensor-Core Kernels in BaseRT}

BaseRT is a C++ inference runtime that targets Metal directly, with no dependency on MLX, PyTorch, CoreML, or any intermediate array framework. Its design (data-driven architecture descriptors, a zero-allocation decode loop, an extensive library of hand-written and fused Metal shaders, and low-overhead command dispatch) is described in prior work~\cite{basert2026}. We add the Metal~4 tensor path as a new family of kernels alongside the existing SIMD kernels, selected at dispatch time.

\subsection{A Tensor-Core Kernel Family for Inference}

We implement cooperative-tensor kernels for the matmul-dominated operators of inference. Each threadgroup owns an output tile and, for each step along the contraction dimension, feeds operand tiles into an MPP \texttt{matmul2d} that accumulates into a register-resident cooperative tensor. The family covers:

\begin{itemize}
  \item \textbf{Dense GEMM} for the attention and feed-forward projection weights, in 4-bit and 8-bit quantisation, reading the same packed-quant weight layout as BaseRT's SIMD GEMM so no reconversion is required.
  \item \textbf{MoE expert GEMM}, a kernel that computes each active expert's projection over its routed tokens.
  \item \textbf{Fused MoE gate/up projection}, which computes the gate and up projections and applies the SiLU activation on the tensor-core accumulators in a single dispatch, removing a global memory round-trip per expert.
  \item \textbf{Prefill attention}, a flash-attention kernel that routes both the QK$^\top$ score product and the PV output product through \texttt{matmul2d} while keeping an exact online softmax (running row maximum, numerically stabilised) between them, so attention output is numerically equivalent to the SIMD FlashAttention path.
\end{itemize}

\subsection{Workload-Aware Dispatch}

The per-chip configuration of BaseRT detects the M5 family at load time. The tensor path is selected for prefill GEMM, MoE expert GEMM, and prefill attention (the compute-bound operators), while decode continues to use BaseRT's specialised GEMV and decode-attention kernels, which are bounded by memory bandwidth and gain nothing from a matmul engine. Dispatch additionally checks shape constraints, falling back to the SIMD kernel whenever they are not met. On any pre-M5 chip, the entire tensor path is bypassed and behaviour is unchanged from the SIMD kernels. The result is a single runtime that transparently exploits the Neural Accelerators of the M5 architecture where they matter and preserves BaseRT's established performance everywhere else.

\section{Evaluation}
\label{sec:eval}

\subsection{Experimental Setup}

\paragraph{Hardware.} All measurements are taken on an Apple M5 Pro with 48\,GB of unified memory, running macOS (Darwin 25.4).

\paragraph{Models.} We benchmark fifteen model configurations across four distinct model families: Qwen3, the newer Qwen3.5/3.6 family (newly supported as of BaseRT v0.1.6), Llama~3.2, and Gemma~4, from sub-1B to 35B parameters. The set spans dense models together with four mixture-of-experts models, namely Gemma-4-26B-A4B, Qwen3-30B-A3B, Qwen3.5-35B-A3B, and Qwen3.6-35B-A3B. Ten configurations use 4-bit quantisation, with 8-bit variants for five of the smaller models (Qwen3-0.6B, Llama-3.2-1B, Llama-3.2-3B, Gemma-4-E2B, and Qwen3.5-2B).

\paragraph{Baselines.} We compare against llama.cpp (build b9960) using its Metal backend with default settings, and mlx-lm 0.31.3 (MLX 0.32.0) with default generation settings. BaseRT (v0.1.6) loads the same checkpoints converted to its native base weight format at matched quantisation. All three engines run on the same M5 Pro device.

\paragraph{Metrics.} We report prompt-processing throughput (pp, tokens/s) at prompt lengths of 128, 256, 512, 1024, and 2048 tokens, and token-generation throughput (tg128, tokens/s) over 128 generated tokens. Each data point is the average of five repetitions.

\subsection{Decode Throughput}
\label{sec:decode}

Table~\ref{tab:decode_m5} reports decode throughput (tg128). Against llama.cpp, BaseRT is faster on all fifteen configurations, by $1.02$--$1.75\times$. The advantage is largest on the smaller dense models, where BaseRT's low-overhead dispatch is a larger fraction of a short per-token budget, and on the 35B-A3B mixture-of-experts models ($1.53$--$1.75\times$), where llama.cpp's measured MoE decode throughput is comparatively low. Against MLX the picture is tighter: BaseRT leads on thirteen of fifteen configurations ($1.02$--$1.33\times$) but trails narrowly on the two large models Gemma-4-26B-A4B and Qwen3.6-27B ($0.98\times$). This is expected, as decode is memory-bandwidth-bound, and the M5's Neural Accelerators do not raise the memory ceiling. Here BaseRT and MLX both operate close to what the hardware's bandwidth allows, and the tensor cores confer no decode advantage. The decode gains that remain come from BaseRT's dispatch and kernel-fusion work, not from the M5 tensor path.

\begin{table}[htbp]
\centering
\caption{Decode throughput (tg128, tok/s) on Apple M5 Pro for BaseRT, llama.cpp, and mlx-lm, grouped into 4-bit and 8-bit blocks. Bold indicates the fastest runtime.}
\label{tab:decode_m5}
\footnotesize
\setlength{\tabcolsep}{5pt}
\begin{tabular}{@{}l r rr rr@{}}
\toprule
 & & \multicolumn{2}{c}{vs.\ llama.cpp} & \multicolumn{2}{c}{vs.\ mlx-lm} \\
\cmidrule(lr){3-4} \cmidrule(lr){5-6}
Model & BaseRT & llama.cpp & Ratio & mlx-lm & Ratio \\
\midrule
\multicolumn{6}{@{}l}{4-bit (Q4)} \\
Qwen3-0.6B        & \textbf{530.8} & 386.1 & 1.37$\times$ & 398.1 & 1.33$\times$ \\
Llama-3.2-1B      & \textbf{342.1} & 267.1 & 1.28$\times$ & 298.0 & 1.15$\times$ \\
Qwen3.5-2B        & \textbf{219.6} & 147.1 & 1.49$\times$ & 208.7 & 1.05$\times$ \\
Llama-3.2-3B      & \textbf{137.0} & 120.0 & 1.14$\times$ & 131.0 & 1.05$\times$ \\
Gemma-4-E2B       & \textbf{149.8} & 103.3 & 1.45$\times$ & 146.9 & 1.02$\times$ \\
Gemma-4-26B-A4B   & 85.0 & 83.3 & 1.02$\times$ & \textbf{87.0} & 0.98$\times$ \\
Qwen3-30B-A3B     & \textbf{105.1} & 96.7 & 1.09$\times$ & 97.7 & 1.08$\times$ \\
Qwen3.5-35B-A3B   & \textbf{110.6} & 72.5 & 1.53$\times$ & 104.8 & 1.06$\times$ \\
Qwen3.6-35B-A3B   & \textbf{110.7} & 63.1 & 1.75$\times$ & 101.9 & 1.09$\times$ \\
Qwen3.6-27B       & 18.1 & 15.4 & 1.18$\times$ & \textbf{18.4} & 0.98$\times$ \\
\midrule
\multicolumn{6}{@{}l}{8-bit (Q8)} \\
Qwen3-0.6B        & \textbf{365.6} & 285.3 & 1.28$\times$ & 295.5 & 1.24$\times$ \\
Llama-3.2-1B      & \textbf{204.4} & 182.8 & 1.12$\times$ & 181.0 & 1.13$\times$ \\
Qwen3.5-2B        & \textbf{132.4} & 110.6 & 1.20$\times$ & 125.9 & 1.05$\times$ \\
Llama-3.2-3B      & \textbf{79.6} & 75.5 & 1.05$\times$ & 75.4 & 1.06$\times$ \\
Gemma-4-E2B       & \textbf{95.9} & 72.7 & 1.32$\times$ & 91.7 & 1.05$\times$ \\
\bottomrule
\end{tabular}
\end{table}

\FloatBarrier
\subsection{Prefill Throughput}
\label{sec:prefill}

The tensor cores transform prompt processing. Tables~\ref{tab:prefill_llama_m5} and~\ref{tab:prefill_mlx_m5} report prefill throughput against llama.cpp and MLX respectively. Against llama.cpp, BaseRT is faster on every configuration at every prompt length, and the margins are large and grow with context: on the dense Gemma-4-E2B the ratio rises from $2.8\times$ at pp128 to $6.3\times$ at pp2048 ($6.4\times$ at 8-bit), and on the MoE Qwen3-30B-A3B from $1.9\times$ to $2.2\times$. Even on the models where llama.cpp is most competitive (Qwen3-0.6B, Qwen3.5-2B), BaseRT leads by $10$--$96\%$. Against MLX, the stronger prefill baseline on Apple Silicon, BaseRT leads on all but three of seventy-five measured points, the exceptions confined to the two smaller Llama models at long context (Llama-3.2-1B and Llama-3.2-3B at pp1024--2048), where MLX edges ahead by $\le3\%$. Everywhere else BaseRT is ahead, by up to $3.9\times$ (Qwen3-30B-A3B at pp128).

Two patterns stand out. First, the widest single margins appear on the mixture-of-experts models: MoE inference is matmul-heavy per active parameter, and BaseRT's tensor-core expert GEMM and fused gate/up kernels route this work onto the M5's Neural Accelerators. Across the four MoE models, BaseRT leads llama.cpp by $29$--$120\%$ and MLX by $15$--$288\%$ over the sweep, with the largest gaps on Gemma-4-26B-A4B and Qwen3-30B-A3B. Second, the advantage generally widens with prompt length against llama.cpp, whose prefill does not scale as well with the growing GEMM, while against MLX it is largest at short prompts and narrows toward long context. BaseRT's lead holds across families and model sizes, from a 0.6B dense model to the 35B mixture-of-experts models.

\begin{table*}[t]
\centering
\caption{Prefill throughput (tok/s) on Apple M5 Pro, BaseRT vs.\ llama.cpp, at prompt lengths 128--2048, grouped into 4-bit and 8-bit blocks. Bold indicates the faster runtime at each prompt length.}
\label{tab:prefill_llama_m5}
\footnotesize
\setlength{\tabcolsep}{3pt}
\begin{tabular}{@{}l rr rr rr rr rr@{}}
\toprule
 & \multicolumn{2}{c}{pp128} & \multicolumn{2}{c}{pp256} & \multicolumn{2}{c}{pp512} & \multicolumn{2}{c}{pp1024} & \multicolumn{2}{c}{pp2048} \\
\cmidrule(lr){2-3} \cmidrule(lr){4-5} \cmidrule(lr){6-7} \cmidrule(lr){8-9} \cmidrule(lr){10-11}
Model & BaseRT & ll.cpp & BaseRT & ll.cpp & BaseRT & ll.cpp & BaseRT & ll.cpp & BaseRT & ll.cpp \\
\midrule
\multicolumn{11}{@{}l}{4-bit (Q4)} \\
Qwen3-0.6B      & \textbf{11{,}873} & 10{,}559 & \textbf{17{,}575} & 13{,}741 & \textbf{20{,}706} & 15{,}509 & \textbf{21{,}213} & 13{,}344 & \textbf{20{,}105} & 10{,}464 \\
Llama-3.2-1B    & \textbf{8{,}936} & 6{,}859 & \textbf{11{,}032} & 9{,}137 & \textbf{12{,}134} & 9{,}740 & \textbf{12{,}084} & 9{,}490 & \textbf{12{,}451} & 8{,}362 \\
Qwen3.5-2B      & \textbf{4{,}594} & 3{,}908 & \textbf{5{,}860} & 4{,}837 & \textbf{6{,}682} & 5{,}297 & \textbf{6{,}914} & 5{,}252 & \textbf{7{,}270} & 5{,}166 \\
Llama-3.2-3B    & \textbf{3{,}663} & 2{,}831 & \textbf{4{,}321} & 3{,}457 & \textbf{4{,}383} & 3{,}665 & \textbf{4{,}407} & 3{,}305 & \textbf{4{,}134} & 3{,}091 \\
Gemma-4-E2B     & \textbf{5{,}780} & 2{,}093 & \textbf{8{,}678} & 2{,}608 & \textbf{13{,}444} & 2{,}742 & \textbf{15{,}363} & 2{,}694 & \textbf{16{,}677} & 2{,}656 \\
Gemma-4-26B-A4B & \textbf{2{,}246} & 1{,}185 & \textbf{2{,}599} & 1{,}587 & \textbf{3{,}307} & 1{,}788 & \textbf{3{,}427} & 1{,}709 & \textbf{3{,}381} & 1{,}618 \\
Qwen3-30B-A3B   & \textbf{2{,}478} & 1{,}280 & \textbf{3{,}199} & 1{,}726 & \textbf{3{,}701} & 2{,}086 & \textbf{3{,}907} & 1{,}964 & \textbf{3{,}834} & 1{,}740 \\
Qwen3.5-35B-A3B & \textbf{1{,}501} & 1{,}065 & \textbf{1{,}974} & 1{,}440 & \textbf{2{,}517} & 1{,}702 & \textbf{2{,}800} & 1{,}653 & \textbf{3{,}102} & 1{,}632 \\
Qwen3.6-35B-A3B & \textbf{1{,}361} & 1{,}042 & \textbf{1{,}831} & 1{,}424 & \textbf{2{,}342} & 1{,}698 & \textbf{2{,}736} & 1{,}640 & \textbf{2{,}788} & 1{,}628 \\
Qwen3.6-27B     & \textbf{440} & 358 & \textbf{476} & 383 & \textbf{494} & 392 & \textbf{498} & 390 & \textbf{497} & 385 \\
\midrule
\multicolumn{11}{@{}l}{8-bit (Q8)} \\
Qwen3-0.6B      & \textbf{11{,}191} & 9{,}695 & \textbf{16{,}926} & 13{,}076 & \textbf{20{,}354} & 14{,}818 & \textbf{21{,}198} & 12{,}921 & \textbf{20{,}078} & 10{,}235 \\
Llama-3.2-1B    & \textbf{8{,}434} & 6{,}739 & \textbf{11{,}392} & 8{,}640 & \textbf{12{,}040} & 9{,}663 & \textbf{11{,}118} & 9{,}123 & \textbf{12{,}434} & 8{,}317 \\
Qwen3.5-2B      & \textbf{4{,}403} & 3{,}987 & \textbf{5{,}795} & 4{,}868 & \textbf{6{,}691} & 5{,}445 & \textbf{6{,}982} & 5{,}400 & \textbf{7{,}268} & 5{,}320 \\
Llama-3.2-3B    & \textbf{3{,}480} & 2{,}738 & \textbf{4{,}237} & 3{,}363 & \textbf{4{,}349} & 3{,}553 & \textbf{4{,}431} & 3{,}367 & \textbf{4{,}389} & 3{,}069 \\
Gemma-4-E2B     & \textbf{4{,}991} & 1{,}992 & \textbf{7{,}825} & 2{,}524 & \textbf{12{,}380} & 2{,}631 & \textbf{14{,}661} & 2{,}600 & \textbf{16{,}264} & 2{,}547 \\
\bottomrule
\end{tabular}
\end{table*}

\begin{table*}[t]
\centering
\caption{Prefill throughput (tok/s) on Apple M5 Pro, BaseRT vs.\ mlx-lm, at prompt lengths 128--2048, grouped into 4-bit and 8-bit blocks. Bold indicates the faster runtime at each prompt length.}
\label{tab:prefill_mlx_m5}
\footnotesize
\setlength{\tabcolsep}{3pt}
\begin{tabular}{@{}l rr rr rr rr rr@{}}
\toprule
 & \multicolumn{2}{c}{pp128} & \multicolumn{2}{c}{pp256} & \multicolumn{2}{c}{pp512} & \multicolumn{2}{c}{pp1024} & \multicolumn{2}{c}{pp2048} \\
\cmidrule(lr){2-3} \cmidrule(lr){4-5} \cmidrule(lr){6-7} \cmidrule(lr){8-9} \cmidrule(lr){10-11}
Model & BaseRT & mlx & BaseRT & mlx & BaseRT & mlx & BaseRT & mlx & BaseRT & mlx \\
\midrule
\multicolumn{11}{@{}l}{4-bit (Q4)} \\
Qwen3-0.6B      & \textbf{11{,}873} & 4{,}600 & \textbf{17{,}575} & 5{,}699 & \textbf{20{,}706} & 12{,}807 & \textbf{21{,}213} & 13{,}732 & \textbf{20{,}105} & 16{,}198 \\
Llama-3.2-1B    & \textbf{8{,}936} & 3{,}987 & \textbf{11{,}032} & 8{,}431 & \textbf{12{,}134} & 9{,}971 & \textbf{12{,}084} & 11{,}490 & 12{,}451 & \textbf{12{,}661} \\
Qwen3.5-2B      & \textbf{4{,}594} & 2{,}328 & \textbf{5{,}860} & 4{,}346 & \textbf{6{,}682} & 5{,}752 & \textbf{6{,}914} & 6{,}644 & \textbf{7{,}270} & 7{,}003 \\
Llama-3.2-3B    & \textbf{3{,}663} & 2{,}531 & \textbf{4{,}321} & 3{,}031 & \textbf{4{,}383} & 3{,}914 & \textbf{4{,}407} & 4{,}270 & 4{,}134 & \textbf{4{,}238} \\
Gemma-4-E2B     & \textbf{5{,}780} & 2{,}634 & \textbf{8{,}678} & 5{,}217 & \textbf{13{,}444} & 9{,}203 & \textbf{15{,}363} & 11{,}645 & \textbf{16{,}677} & 13{,}187 \\
Gemma-4-26B-A4B & \textbf{2{,}246} & 806 & \textbf{2{,}599} & 1{,}322 & \textbf{3{,}307} & 1{,}783 & \textbf{3{,}427} & 2{,}247 & \textbf{3{,}381} & 2{,}419 \\
Qwen3-30B-A3B   & \textbf{2{,}478} & 639 & \textbf{3{,}199} & 1{,}090 & \textbf{3{,}701} & 1{,}548 & \textbf{3{,}907} & 2{,}267 & \textbf{3{,}834} & 2{,}676 \\
Qwen3.5-35B-A3B & \textbf{1{,}501} & 602 & \textbf{1{,}974} & 1{,}112 & \textbf{2{,}517} & 1{,}509 & \textbf{2{,}800} & 2{,}044 & \textbf{3{,}102} & 2{,}427 \\
Qwen3.6-35B-A3B & \textbf{1{,}361} & 596 & \textbf{1{,}831} & 1{,}122 & \textbf{2{,}342} & 1{,}472 & \textbf{2{,}736} & 1{,}988 & \textbf{2{,}788} & 2{,}426 \\
Qwen3.6-27B     & \textbf{440} & 305 & \textbf{476} & 379 & \textbf{494} & 442 & \textbf{498} & 475 & \textbf{497} & 486 \\
\midrule
\multicolumn{11}{@{}l}{8-bit (Q8)} \\
Qwen3-0.6B      & \textbf{11{,}191} & 5{,}134 & \textbf{16{,}926} & 7{,}080 & \textbf{20{,}354} & 13{,}146 & \textbf{21{,}198} & 17{,}463 & \textbf{20{,}078} & 16{,}037 \\
Llama-3.2-1B    & \textbf{8{,}434} & 4{,}059 & \textbf{11{,}392} & 7{,}874 & \textbf{12{,}040} & 8{,}872 & 11{,}118 & \textbf{11{,}376} & \textbf{12{,}434} & 12{,}101 \\
Qwen3.5-2B      & \textbf{4{,}403} & 2{,}140 & \textbf{5{,}795} & 3{,}909 & \textbf{6{,}691} & 5{,}168 & \textbf{6{,}982} & 6{,}228 & \textbf{7{,}268} & 6{,}734 \\
Llama-3.2-3B    & \textbf{3{,}480} & 2{,}035 & \textbf{4{,}237} & 2{,}640 & \textbf{4{,}349} & 3{,}482 & \textbf{4{,}431} & 3{,}971 & \textbf{4{,}389} & 4{,}145 \\
Gemma-4-E2B     & \textbf{4{,}991} & 2{,}398 & \textbf{7{,}825} & 4{,}788 & \textbf{12{,}380} & 8{,}274 & \textbf{14{,}661} & 10{,}752 & \textbf{16{,}264} & 12{,}355 \\
\bottomrule
\end{tabular}
\end{table*}

\FloatBarrier
\subsection{Discussion of Results}

Sections \ref{sec:decode} and \ref{sec:prefill} present consistent results. Decode is memory-bandwidth-bound: on the M5 Pro all three engines cluster near the bandwidth ceiling on the larger models, BaseRT's remaining decode advantage comes from dispatch and fusion rather than the tensor cores, and MLX occasionally edges ahead. Prefill is compute-bound, and here the M5's Neural Accelerators are decisive: by routing prompt GEMM, MoE expert GEMM, and prefill attention through the Metal~4 tensor path, BaseRT converts the M5's matmul throughput into the highest prompt-processing throughput measured on this hardware, with the largest gains on the matmul-heavy mixture-of-experts models. Because time-to-first-token is dominated by prefill, these gains translate directly into lower interactive latency for long-prompt and agentic workloads, exactly the use cases that motivate on-device inference.

\section{Discussion}
\subsection{Limitations}
\label{sec:limitations}

The tensor-core kernels apply to the compute-bound phases of inference. They do not raise the decode throughput ceiling set by memory bandwidth. Realising the prefill gains requires M5-family hardware and a Metal~4 toolchain. On earlier chips BaseRT falls back to its SIMD kernels with unchanged behaviour. As in prior work, BaseRT targets single-device, single-user inference: it does not implement continuous batching, multi-request parallel decoding, or tensor parallelism, and it targets Metal exclusively. The evaluation is conducted on a single M5 Pro device. The behaviour on the M5 base and Max tiers, which differ in core count and memory bandwidth, is left for future measurement.

\subsection{Future Work}

Decode remains memory-bound, but speculative decoding turns part of the decode workload into a batched verification step whose matmuls are compute-bound and therefore amenable to the tensor-core path. Combining speculative decoding with the M5 Neural Accelerators is a promising direction for raising decode throughput indirectly. The prefill advantage also invites revisiting the balance between prefill and decode in scheduling long-context and multimodal workloads.

\section{Conclusion}

We presented an extension of BaseRT that targets the Apple M5's per-core GPU Neural Accelerators through the Metal~4 tensor API. By adding hand-written cooperative-tensor kernels for dense and mixture-of-experts GEMM, fused expert projections, and prefill attention, and by applying them only to the compute-bound phases of inference, BaseRT turns the M5's Neural Accelerators into best-in-class inference throughput on this hardware. On the Apple M5 Pro, this yields up to $6.4\times$ higher prefill throughput than llama.cpp and $3.9\times$ higher than MLX, alongside decode gains bounded, as expected, by memory bandwidth. These results push the performance ceiling of on-device LLM inference on Apple Silicon markedly higher and demonstrate that exploiting the M5's tensor cores translates into substantial, measurable throughput gains for on-device inference. BaseRT is available at \url{https://github.com/basecompute/baseRT}.

\bibliographystyle{unsrtnat}
\bibliography{references}

@article{benazir2025profiling,
  author       = {Benazir, Afsara and Lin, Felix Xiaozhu},
  title        = {{Profiling Large Language Model Inference on Apple Silicon:
                  A Quantization Perspective}},
  journal      = {arXiv preprint arXiv:2508.08531},
  year         = {2025},
  eprint       = {2508.08531},
  archivePrefix = {arXiv}
}

@techreport{stanfordaiindex2025,
  author       = {Maslej, Nestor and others},
  title        = {{Artificial Intelligence Index Report 2025}},
  institution  = {Stanford Institute for Human-Centered Artificial Intelligence (HAI)},
  year         = {2025},
  note         = {arXiv:2504.07139}
}

@inproceedings{jang2025edgefirst,
  author       = {Jang, SiYoung and Morabito, Roberto},
  title        = {{Edge-First Language Model Inference: Models, Metrics, and Tradeoffs}},
  booktitle    = {45th IEEE International Conference on Distributed Computing Systems (ICDCS)},
  year         = {2025},
  eprint       = {2505.16508},
  archivePrefix = {arXiv},
  url          = {https://arxiv.org/abs/2505.16508}
}

@misc{mckinsey2025neweconomics,
  author       = {{McKinsey \& Company}},
  title        = {{The New Economics of Enterprise Technology in an AI World}},
  year         = {2025},
  month        = {May},
  howpublished = {\url{https://www.mckinsey.com/capabilities/tech-and-ai/our-insights/}}
}

@inproceedings{huang2025corki,
  author       = {Huang, Xuan and others},
  title        = {{DaDu-Corki: Algorithm-Architecture Co-Design for
                  Embodied AI-powered Robotic Manipulation}},
  booktitle    = {Proceedings of the 52nd Annual International Symposium
                  on Computer Architecture (ISCA)},
  year         = {2025},
  eprint       = {2407.04292},
  archivePrefix = {arXiv},
  url          = {https://arxiv.org/abs/2407.04292}
}

@article{masterman2026agentic,
  author       = {Masterman, Tula and others},
  title        = {{Agentic Artificial Intelligence: Architectures,
                  Taxonomies, and Evaluation of Large Language Model Agents}},
  journal      = {arXiv preprint arXiv:2601.12560},
  year         = {2026},
  eprint       = {2601.12560},
  archivePrefix = {arXiv},
  url          = {https://arxiv.org/abs/2601.12560}
}

@misc{anthropicstatus2026,
  author       = {{Anthropic}},
  title        = {{Anthropic Status Page: Incident History}},
  year         = {2026},
  howpublished = {\url{https://status.anthropic.com}},
  note         = {Accessed: April 2026. Documented 167 incidents
                  between October 2025 and April 2026}
}

@misc{openaistatus2026,
  author       = {{OpenAI}},
  title        = {{OpenAI Status Page: Incident History}},
  year         = {2026},
  howpublished = {\url{https://status.openai.com}},
  note         = {Accessed: April 2026}
}

@misc{llamacpp,
  author       = {Gerganov, Georgi and others},
  title        = {{llama.cpp}: {LLM} Inference in {C/C++}},
  year         = {2023},
  howpublished = {\url{https://github.com/ggml-org/llama.cpp}}
}

@misc{mlx,
  author       = {Hannun, Awni and Digani, Jagrit and Katharopoulos, Angelos and Collobert, Ronan},
  title        = {{MLX}: Efficient and Flexible Machine Learning on {Apple} Silicon},
  year         = {2023},
  howpublished = {\url{https://github.com/ml-explore/mlx}}
}

@article{basert2026,
  author        = {Rathnayaka, Prabod and Waschkowski, Fabian and Wesemann, Lukas},
  title         = {{BaseRT: Best-in-Class LLM Inference on Apple Silicon via Native Metal}},
  journal       = {arXiv preprint arXiv:2607.00501},
  year          = {2026},
  eprint        = {2607.00501},
  archivePrefix = {arXiv},
  url           = {https://arxiv.org/abs/2607.00501}
}

@techreport{mozilla2026opensource,
  author       = {{Mozilla Foundation}},
  title        = {{State of Open Source AI 2026}},
  institution  = {Mozilla Foundation},
  year         = {2026},
  url          = {https://stateofopensource.ai/state-of-open-source-ai-2026.pdf},
  note         = {Accessed: July 2026}
}

@misc{apple2025m5,
  author       = {{Apple}},
  title        = {{Apple unleashes M5, the next big leap in AI performance for Apple silicon}},
  year         = {2025},
  month        = oct,
  howpublished = {Apple Newsroom},
  url          = {https://www.apple.com/newsroom/2025/10/apple-unleashes-m5-the-next-big-leap-in-ai-performance-for-apple-silicon/},
  note         = {Accessed: July 2026}
}

@misc{apple_metal4,
  author       = {{Apple}},
  title        = {{Metal 4}},
  year         = {2025},
  howpublished = {Apple Developer Documentation},
  url          = {https://developer.apple.com/documentation/metal/metal-4},
  note         = {Accessed: July 2026}
}

@misc{apple_mpp,
  author       = {{Apple}},
  title        = {{Metal Performance Primitives (MPP) Programming Guide}},
  year         = {2026},
  month        = mar,
  howpublished = {Apple Inc., Version 1},
  note         = {Accessed: July 2026}
}

@misc{apple_msl,
  author       = {{Apple}},
  title        = {{Metal Shading Language Specification}},
  year         = {2026},
  howpublished = {Apple Inc., Version 4.1},
  note         = {Accessed: July 2026}
}

@misc{apple_wwdc25_tensor,
  author       = {{Apple}},
  title        = {Discover {Metal} 4 Tensors and {Metal} Performance Primitives},
  howpublished = {Apple Worldwide Developers Conference (WWDC)},
  year         = {2025},
  url          = {https://developer.apple.com/videos/play/wwdc2025/}
}

@misc{mckinsey2025aiworkloads,
  author       = {Arora, Chhavi and Sorel, Marc and Sachdeva, Pankaj and others},
  title        = {The Next Big Shifts in {AI} Workloads and Hyperscaler Strategies},
  year         = {2025},
  month        = dec,
  howpublished = {McKinsey \& Company},
  url          = {https://www.mckinsey.com/industries/technology-media-and-telecommunications/our-insights/the-next-big-shifts-in-ai-workloads-and-hyperscaler-strategies},
  note         = {Accessed: April 2026}
}

@misc{gartner2025crossborder,
  author       = {Fritsch, Joerg and {Gartner, Inc.}},
  title        = {Gartner Predicts 40\% of {AI} Data Breaches Will Arise
                  from Cross-Border {GenAI} Misuse by 2027},
  year         = {2025},
  month        = feb,
  howpublished = {Gartner Newsroom},
  url          = {https://www.gartner.com/en/newsroom/press-releases/2025-02-17-gartner-predicts-forty-percent-of-ai-data-breaches-will-arise-from-cross-border-genai-misuse-by-2027},
  note         = {Accessed: April 2026}
}

@techreport{nist2024genai,
  author      = {{National Institute of Standards and Technology}},
  title       = {Artificial Intelligence Risk Management Framework:
                 Generative Artificial Intelligence Profile},
  number      = {NIST AI 600-1},
  institution = {U.S. Department of Commerce},
  year        = {2024},
  month       = jul,
  url         = {https://nvlpubs.nist.gov/nistpubs/ai/NIST.AI.600-1.pdf}
}

\end{document}